\documentclass{iopjournal}

\usepackage{multirow}
\usepackage{amsmath}
\usepackage{tikz}
\usetikzlibrary{positioning,arrows.meta}
\usepackage{subcaption}

\renewcommand{\iopjournalshort}{Mach. Learn.: Sci. Technol.}
\renewcommand{\iopvolume}{7}
\renewcommand{\iopyear}{2026}
\renewcommand{\ioparticleid}{055024}
\renewcommand{\iopdoi}{10.1088/2632-2153/aea687}
\renewcommand{\iopreceived}{17 March 2026}
\renewcommand{\ioprevised}{19 August 2026}
\renewcommand{\iopaccepted}{11 September 2026}
\renewcommand{\ioppublished}{24 September 2026}

\renewcommand{\ioplicense}{Creative Commons Attribution 4.0 licence}
\renewcommand{\ioprunningauthor}{I Xiotidis {\it et al}\ }

\begin{document}

\articletype{Paper} 

\title{AMD Versal AI-Engines for fixed latency environments}

\author{I. Xiotidis$^{1,*}$\orcid{0000-0003-1401-4748}, 
N. Clarke Hall$^{1,2}$\orcid{0000-0001-9236-7325}, 
T. Du$^{1,3}$, 
N. Konstantinidis$^2$\orcid{0000-0002-4140-6360} and 
D.~W. Miller$^{3,4}$\orcid{0000-0002-9485-9435}}

\affil{$^1$CERN, Geneva, Switzerland}

\affil{$^2$Department of Physics and Astronomy, UCL, London, United Kingdom}

\affil{$^3$Department of Physics, Enrico Fermi Institute, University of Chicago, Chicago, IL, United States of America}

\affil{$^4$Kavli Institute for Cosmological Physics, University of Chicago, Chicago, IL, United States of America}

\affil{$^*$Author to whom any correspondence should be addressed.}

\email{ioannis.xiotidis@cern.ch}


\begin{abstract}

Complex, high-throughput data acquisition and processing systems, such as those used in high-energy physics experiments, are increasingly moving sophisticated pattern recognition and data compression algorithms closer to the sensors themselves. To meet these needs, programmable device manufacturers offer multi-silicon die packages that commonly include dedicated co-processors within the same package. We present a technical study of a new family of such co-processors from AMD Xilinx, the Adaptive Intelligence (AI) Engine, or AIE, as part of the Versal\textsuperscript{\texttrademark} architecture. Specifically, we focus on the deployment capabilities of AIEs in fixed latency environments with latency constraints of the order of microseconds, such as those typically encountered in colliding beam experiments like at CERN's Large Hadron Collider. We evaluate the performance of a vectorised implementation of both a Boosted Decision Tree and a Convolutional Neural Network, thereby demonstrating the feasibility of deploying AIEs for Machine Learning applications in such environments and their use as possible alternatives to traditional programmable logic-based implementations. In addition we are including an example of a CNN based architecture prototyped for the ATLAS Phase-II trigger, evaluated on open-data. We conclude that AIEs can augment the compute capabilities of field programmable gate array-based processing systems with microsecond-latency requirements.

\end{abstract}

\section{Introduction}

Nearly every arena of modern technology is witnessing the need for compute and data-intensive applications to be co-located with the sensors that are generating those data. 
From telecommunications, to medicine, to robotics, and indeed to large-scale scientific experiments, the appetite for so-called ``edge-computing'' for complex, yet power-efficient, compute capabilities has skyrocketed~\cite{edge_comp_1, edge_comp_2}. In the coming decade, for example, the Large Hadron Collider (LHC) will enter its High-Luminosity (HL) era and will bring with it exponential increases in a number of data processing metrics~\cite{hllhc_1, hllhc_2}. The instrumentation systems in experiments such as ATLAS and CMS~\cite{atlas_ref, cms_ref}, tasked with selecting, processing, and recording these data for permanent storage and analysis, aim to cope with the increase in the complexity of the data generated in these collisions by upgrading and adopting novel technologies for their Trigger and Data Acquisition (TDAQ) systems~\cite{atlas_phase_II, cms_phase_II}. The focus of this work is on evaluating the technical capabilities of deploying a new class of high-performance edge-computing devices in the hard-real-time fixed-latency layers of these upgraded TDAQ architectures.

\begin{figure}[h]
    \centering
    \includegraphics[width=0.4\linewidth]{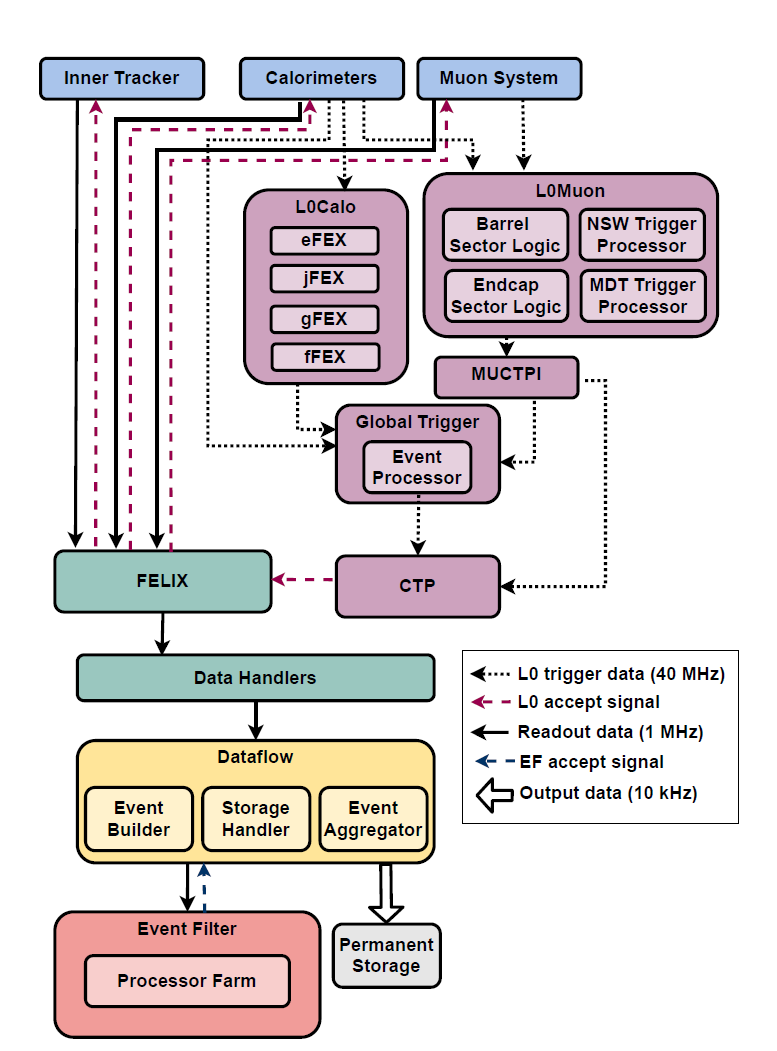}
    \caption{The ATLAS TDAQ system for the HL-LHC upgrade. The lilac boxes show the Level-0 hardware trigger implemented in custom hardware boards. The green and yellow boxes show the data streaming infrastructure based on custom hardware cards and commodity servers. Finally, with salmon the Event Filter processing farm based on commodity servers and potentially GPUs or FPGAs~\cite{atlas_phase_II}.}
    \label{fig:ATLASTDAQPhaseII}
\end{figure}

We focus on the upgraded TDAQ architecture planned for the ATLAS experiment at the LHC. as an example of such a system. Its design follows a robust hybrid approach that combines a custom, hardware-based, fixed-latency compute layer, followed by a commodity compute architecture layer that resembles a modern data centre, as seen in Fig. \ref{fig:ATLASTDAQPhaseII}. The first stage hardware layer (Level-0) will be implemented in all-custom electronics, based primarily on Field Programmable Gate Arrays (FPGAs), whereas the second layer uses a computer farm (Event Filter) which might include Graphics Processing Units (GPUs) or FPGA-based compute accelerators. The Level-0 trigger system is designed to receive the detector input at a rate of 40~MHz and output the events that pass coarse selections at a rate of 1~MHz, with a fixed maximum latency of 10~$\mu$s. Subsequently, the Event Filter farm will apply more elaborate selections on the 1~MHz input stream from the Level-0 Trigger to further reduce to approximately 10~kHz the rate of events selected for permanent storage and offline analysis. 

A new feature of the upgraded ATLAS Level-0 TDAQ system concerns the implementation of a Global Trigger~\cite{atlas_global_trigger}. The ATLAS Global Trigger is a time-multiplexed system which will receive the full-granularity information from multiple subsystems of the experiment ($\mathcal{O}(50)$~Tbps) allowing trigger algorithms to run on the full-event information. This represents a significant evolution compared to the current ATLAS hardware trigger system, which makes data selection decisions based on lower-granularity information. Due to the large number of high-bandwidth input data links and the complexity of the selection algorithms, the Global Trigger is designed to use the latest generation of AMD Xilinx Versal\textsuperscript{\texttrademark} Premium FPGAs~\cite{VP_1, VP_2}. 

\subsection{AMD Versal Premium FPGAs}

The new Versal family of chips from AMD Xilinx follows a paradigm shift with respect to traditional FPGA architectures. The Versal devices are heterogeneous multi-die devices aiming to offload specific tasks to dedicated chiplets, which are optimised for those tasks (e.g. decoding of Ethernet packets, digital signal processing, etc.). This paradigm shift reimagines the FPGA fabric as a set of heterogeneous processing devices within the same package. The communication between the sub-components occurs either via dedicated fixed latency pins or via the Network-on-Chip (NoC)~\cite{NoC}.

A new addition to the list of available co-processor units is a dedicated processor optimised for low-latency arithmetic, called Adaptive Intelligence (AI) Engines~\cite{AIE_1}. These AIE co-processor systems are constructed as 2D arrays consisting of multiple individual AIE \textit{tiles} that provide a scalable co-processor package, ranging from 10s to 100s of AIEs in a single device. The 2D array architecture introduces constraints on the total bandwidth available to each tile, as tiles in the first row closer to the input data links received full bandwidth inputs. 

\begin{figure} [h]
    \centering
    \includegraphics[width=0.7\linewidth]{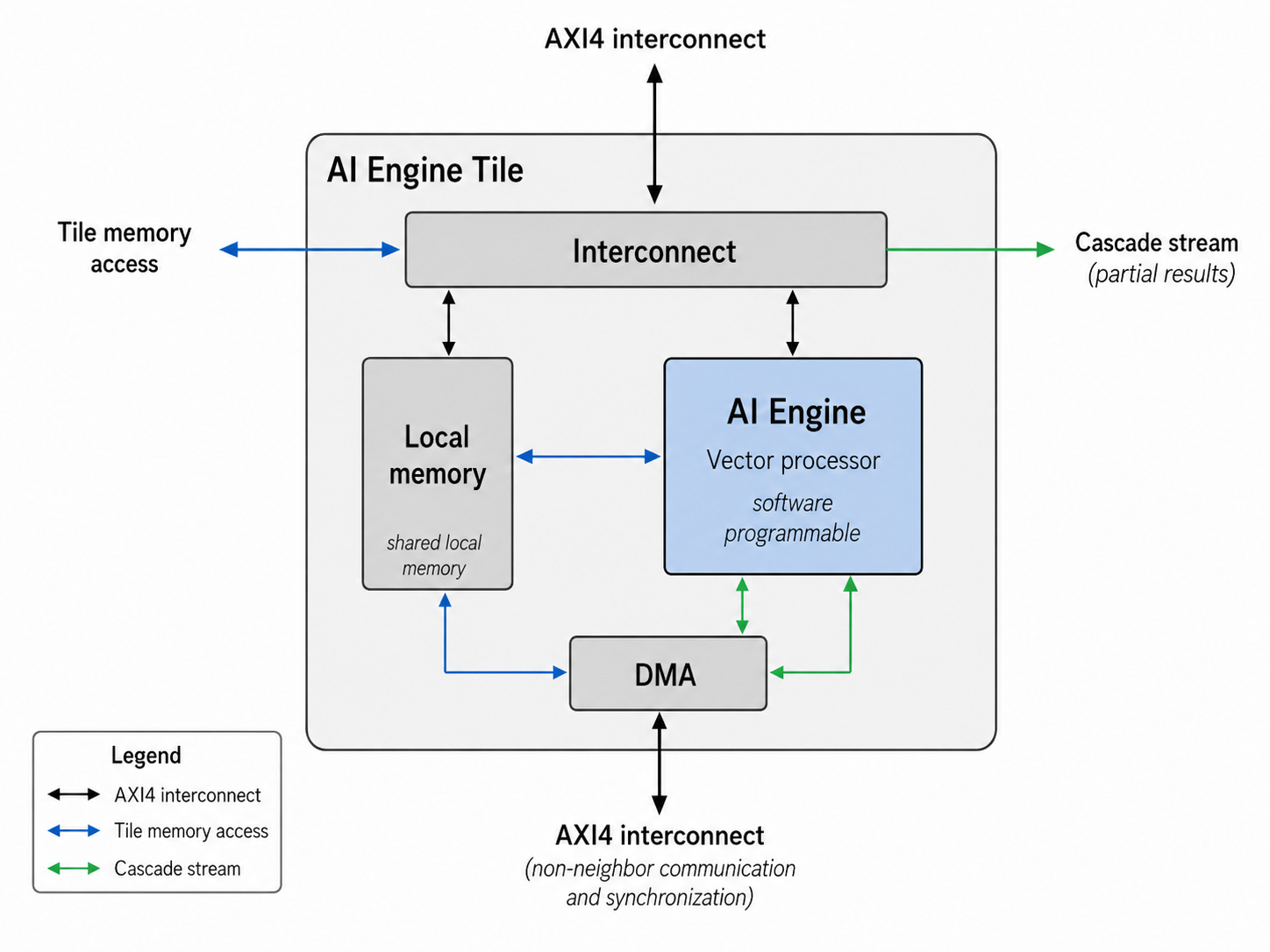}
    \caption{Simplified AI Engine tile sub-components where the 32~kB memory is indicated with the grey box, the input/output components are on the left (streaming or DMA based), the two processing units (vector, scalar) are indicated with the light grey box.~\cite{AIE_1}}
    \label{fig:AIComponents}
\end{figure}

Currently, AMD Xilinx has released three versions of AIEs integrated in different Versal device packages. The primary distinguishing characteristics between the AIE versions are the supported types (e.g. float16) and the available memory on the tile. A summary of the various features of the different AIE versions is provided in Tab.~\ref{tab:AIEVersions}.  

\begin{table}[h]
    \centering
    \begin{tabular}{c c}
        \hline
        \textbf{AIE Version} & \textbf{Feature} \\
        \hline
        AIE-v1.0 & 32kB of memory, 1GHz clock \\
        AIE-ML & 64kB of memory, 1.15GHz clock \\
        AIE-ML v2.0 & 64kB of memory, 1.15GHz, float16 support \\
        \hline
    \end{tabular}
    \caption{Specifications between the different AMD AI Engine versions.}
    \label{tab:AIEVersions}
\end{table}

\section{Model implementations}

Following the introduction of the AI Engines architecture, several studies have investigated its performance in high-throughput applications~\cite{throughput_1, throughput_2}. More recent work has focused on integrating AI Engine kernels with established machine-learning (ML) deployment frameworks, such as hls4ml~\cite{hls4ml}, as demonstrated by AIE4ML~\cite{aie4ml}. These developments provide valuable tools and reference designs for deploying ML algorithms on heterogeneous Versal devices. The present study instead focuses specially on the implementation of ML inference under the microsecond-scale, fixed-latency constraints characteristic of the hardware trigger systems envisioned for the HL-LHC.

In such environments, the large input bandwidth and strict timing requirements restrict the choice of suitable devices to those providing both substantial computational capacity and large numbers of multi-gigabit transceivers. The implementations presented here therefore target the first-generation AI Engine architecture in the Versal Premium device family. In addition to its vector-processing capabilities, this architecture supports single-precision floating-point arithmetic, enabling trained models to be deployed using float32 inputs, parameters and intermediate calculations without first requiring integer quantisation. The results below also discuss the comparison with the integer only implementations.

Three classes of ML operations are considered. First a Boosted Decision Tree (BDT) inference kernel is implemented using bit-mask-based QuickScorer~\cite{QuickScorer} traversal designed to reduce the dependence on conventional branch-by-branch tree evaluation. Second, a streaming two-dimensional convolution kernel which uses line-buffers to construct convolution windows and exploit the vector multiply-accumulate (MAC) capabilities of the AIE. Finally, a Dense-layer kernel implementing the matrix-vector operations required by the classification stages of complete neural-network models. Together, the convolutional and Dense kernels allow complete CNN inference pipelines to be constructed within the AIE array, while the BDT kernel provides an alternative inference architecture for models trained on tabular physics observables. The following sections describe the mapping and implementation of each operation.

\subsection{Boosted Decision Tree kernel}

BDTs are widely used in particle physics as they are ML-based classifiers combining weak decision trees into a single, strong predictor~\cite{bdt_1, bdt_2, bdt_3}. For an ensemble containing (T) trees, the output implemented by the kernels is.

\begin{equation} \label{eq:BDT}
    y = b + \alpha \sum_{t=0}^{T-1}v_{t,\ell_t(\mathbf{x})},
\end{equation}

where (b) is the model base value, ($v_{t,\ell(\mathbf{x})}$) is the float32 value stored in the leaf selected by input ($\mathbf{x}$), and ($\alpha=1$) or ($1/T$) for sum or average aggregation, respectively. Boosting coefficients and learning-rate factors are absorbed into the stored leaf values during model conversion. No sigmoid, softmax or other output transformation is applied within the kernel.

Conventional tree inference follows one branch at each internal node until reaching a leaf. This introduces a sequence of data-dependent control-flow operations that is not naturally mapped onto the vector processor of an AIE. The implementation presented here instead adopts the bit-vector representation introduced by the QuickScorer algorithm. For a tree with ($L_t$) leaves, a 32b candidate mask is initialised as:

\begin{equation} \label{eq:QuickScorerEq1}
    M_t^{(0)} = 2^{L_t}-1,
\end{equation}

such that every leaf is initially a possible output. Each internal-node condition has the form ($x_f\leq\theta$). The complementary result, ($x_f>\theta$), is associated with an exit mask whose zero-valued bits identify leaves that are no longer reachable. Applicable masks are combined using bitwise AND operations, leaving the bit associated with the selected leaf set. Leaves are ordered from left to right, and the output leaf is obtained from the least-significant set bit of the final candidate mask.

To avoid separately applying every node mask, the implementation groups split thresholds by feature and precomputes their cumulative, or prefix, mask states. For feature (f), the ordered thresholds are compared with the input value to determine a threshold bucket:

\begin{equation} \label{eq:QuickScorerEq2}
    q_f = \sum_{j=0}^{N_f-1}\mathbf{1}!\left(x_f>\theta_{f,j}\right).
\end{equation}

The corresponding prefix state ($P_{f,q_f,t}$) is then applied to every tree:

\begin{equation} \label{eq:QuickScorerEq3}
    M_t\leftarrow M_t\wedge P_{f,q_f,t}
\end{equation}

\begin{figure}[h]
    \centering
    \includegraphics[width=0.75\linewidth]{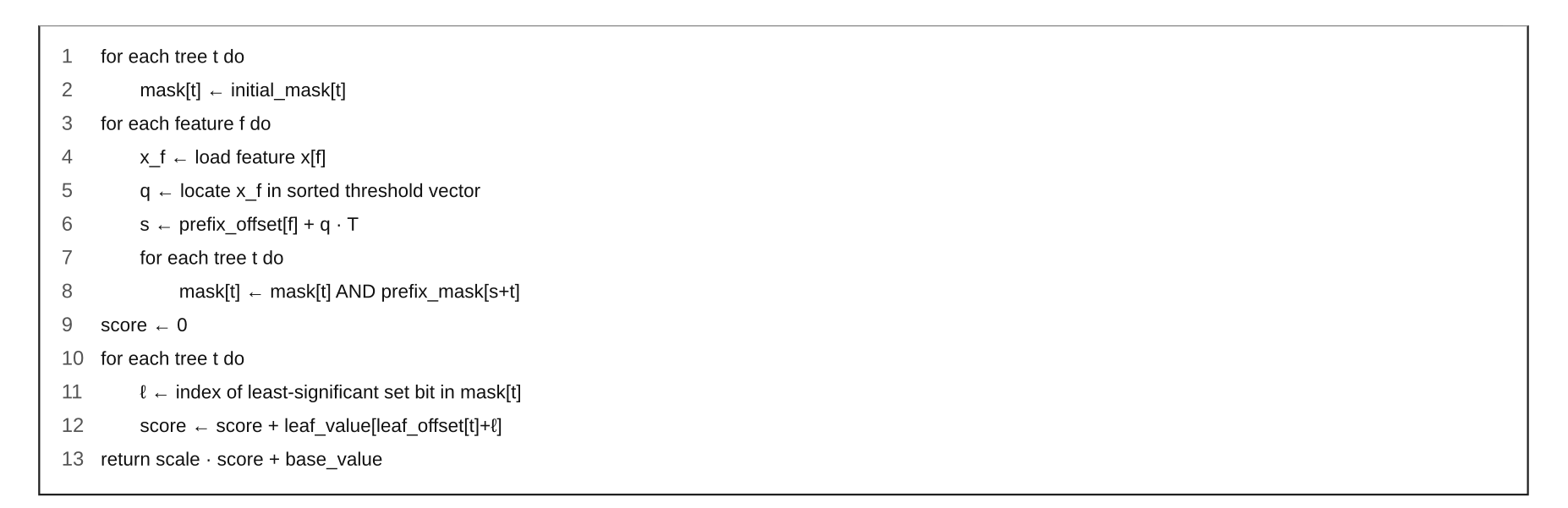}
    \caption{QuickScorer prefix inference for a single input event. Prefix masks encode the cumulative leaf eliminations associated with each feature and threshold bucket. When the forest is divided across multiple workers, the input features are broadcast, and the resulting float32 partial scores are combined through a binary AIE reduction tree.}
    \label{fig:BDTAlgorithm1}
\end{figure}

After all input features have been processed, the least-significant set bit of ($M_t$) identifies ($\ell_t(\mathbf{x})$), and its float32 leaf value is accumulated into the ensemble score. The complete procedure is summarised in Fig. \ref{fig:BDTAlgorithm1}.

\begin{figure}[h]
    \centering
    \includegraphics[width=0.75\linewidth]{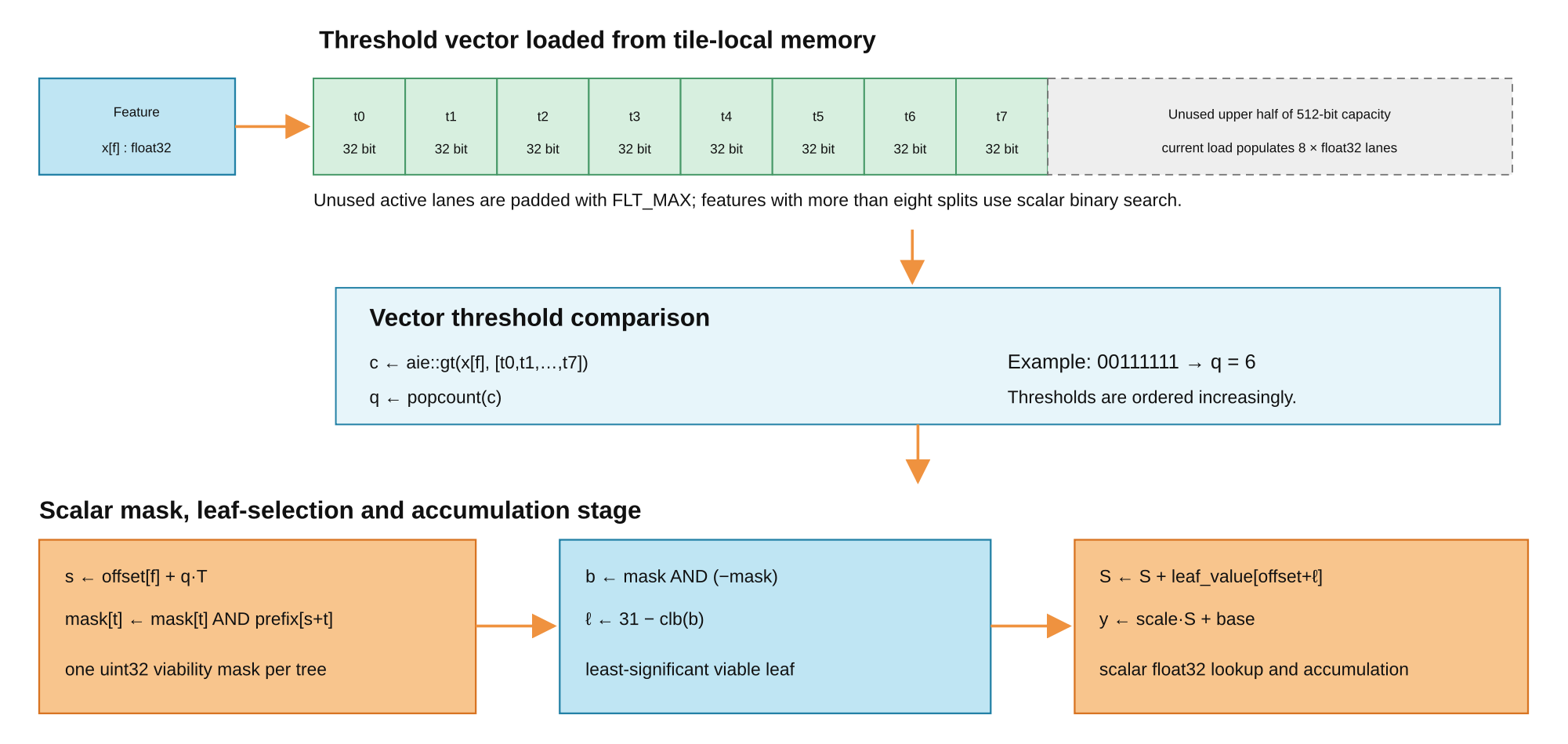}
    \caption{Mapping of the float32 QuickScorer prefix kernel onto the AIE processing units. Eight ordered thresholds are evaluated concurrently during vectorised bucket selection. Prefix-mask updates, leaf resolution and score accumulation are performed by the scalar processor.}
    \label{fig:BDTImage}
\end{figure}

The current implementation vectorises threshold-bucket selection. Up to eight ordered float32 thresholds are loaded into a 256b vector and compared concurrently with a scalar input feature, as illustrated in Fig. \ref{fig:BDTImage}. Unused lanes are padded; when a feature has more than eight split thresholds, a scalar binary search is used. The subsequent prefix-mask updates, leaf identification and float32 score accumulation are performed using scalar instructions. The figure therefore distinguishes the vectorised comparison stage from the scalar mask and accumulation stages rather than implying utilisation of all sixteen float32 lanes available in the conceptual 512b vector capacity.

The model is allowed to contain unbalanced trees with different depths. The QuickScorer representation supports up to 32 leaves per tree, corresponding to a maximum of depth 5. Models with depths up to eight can instead use the direct-traversal implementation. The limit can be enforced during training through the maximum-depth hyperparameter, avoiding an a posteriori transformation of the trained trees into concatenated-observable lookup tables. Thresholds, prefix masks, leaf values and offsets are stored in separate aligned regions of tile-local memory.

When the complete forest does not fit efficiency within one worker, contiguous groups of trees are assigned to multiple AIE workers. The input feature stream is replicated through a binary tree of stream-connected broadcast kernels. Each worker produces one float32 partial sum, and these values are combined using binary tree of AIE reduction kernels. The averaging factor, when required, and the base value are applied only once after the final reduction. Consequently, the design scales the ensemble across AIE tiles without requiring a programmable-logic adder tree.

\subsection{2D Convolution kernel}

A two-dimensional convolution calculates an output feature map (Y) from an input tensor (X) and a set of learned filters (W). Similarly to BDTs CNNs are widely used in HEP experiments~\cite{cnn_1, cnn_2}. The operations implemented by the kernel is

\begin{equation} \label{eq:CNNEquation1}
    Y_{o_h, o_w, o_c} = \sigma\left[b_{o_c}+\sum_{i_c=0}^{C_{\mathrm{in}}-1}\sum_{k_h=0}^{K_h-1}\sum_{k_w=0}^{K_w-1}W_{k_h, k_w, i_c, _c}X_{o_hS_h+k_h,;o_wS_w+k_w,;i_c}\right],
\end{equation}

where ($S_h$) and ($S_w$) denote the vertical and horizontal strides, ($b_{o_c}$) is the bias associated with output channel ($o_c$), and ($\sigma$) is an optional activation. The implementation considered here uses float32 inputs, weights, biases and outputs, with 8-element accumulators. The optimised kernel supports compile-time kernel and spatial dimensions, unit dilation, a single input channel per group and up to eight output channels per vector block. Both VALID convolution and a zero-padded variant are supported, although the following description focuses on the branch-free VALID case.

\begin{figure}[h]
    \centering
    \includegraphics[width=0.75\linewidth]{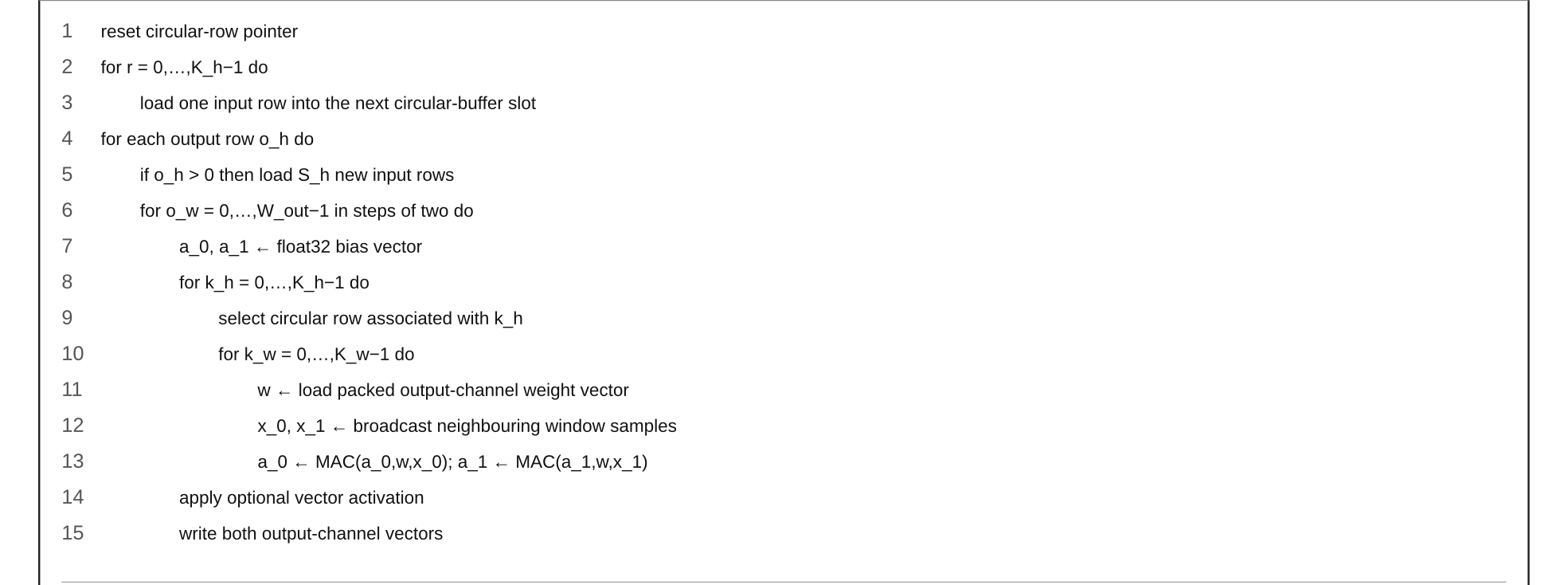}
    \caption{Streaming execution of the float32 Conv2D kernel. A circular $K_h$-row buffer is filled from the row-major input stream, after which pairs of neighbouring output positions are evaluated using two independent vector accumulator chains.}
    \label{fig:CNNAlgorithm2}
\end{figure}

Input samples arrive in row-major order and are stored in a circular line buffer containing ($K_h$) complete rows. After the initial ($K_h$)-row prologue, the first output row can be evaluated. Between successive output row, ($S_h$) new rows are received and written into the slots containing the oldest buffered data. The convolution window is not copied into a flattened ($K_h\times K_w$) patch. Instead, each kernel row selects the corresponding circular-buffer row, and the required samples are addressed directly using the output-column, stride and kernel-column indices. Consequently, the storage requirement of the line buffer scales as approximately

\begin{equation} \label{eq:CNNEquation2}
    B_{\mathrm{line}} = 4 K_hW_{\mathrm{in}}\ \mathrm{bytes}
\end{equation}

for the single-channel float32 implementation, rather than with the complete input feature map. The streaming sequence is summarised in Fig. \ref{fig:CNNAlgorithm2}.

\begin{figure}[h]
    \centering
    \includegraphics[width=0.75\linewidth]{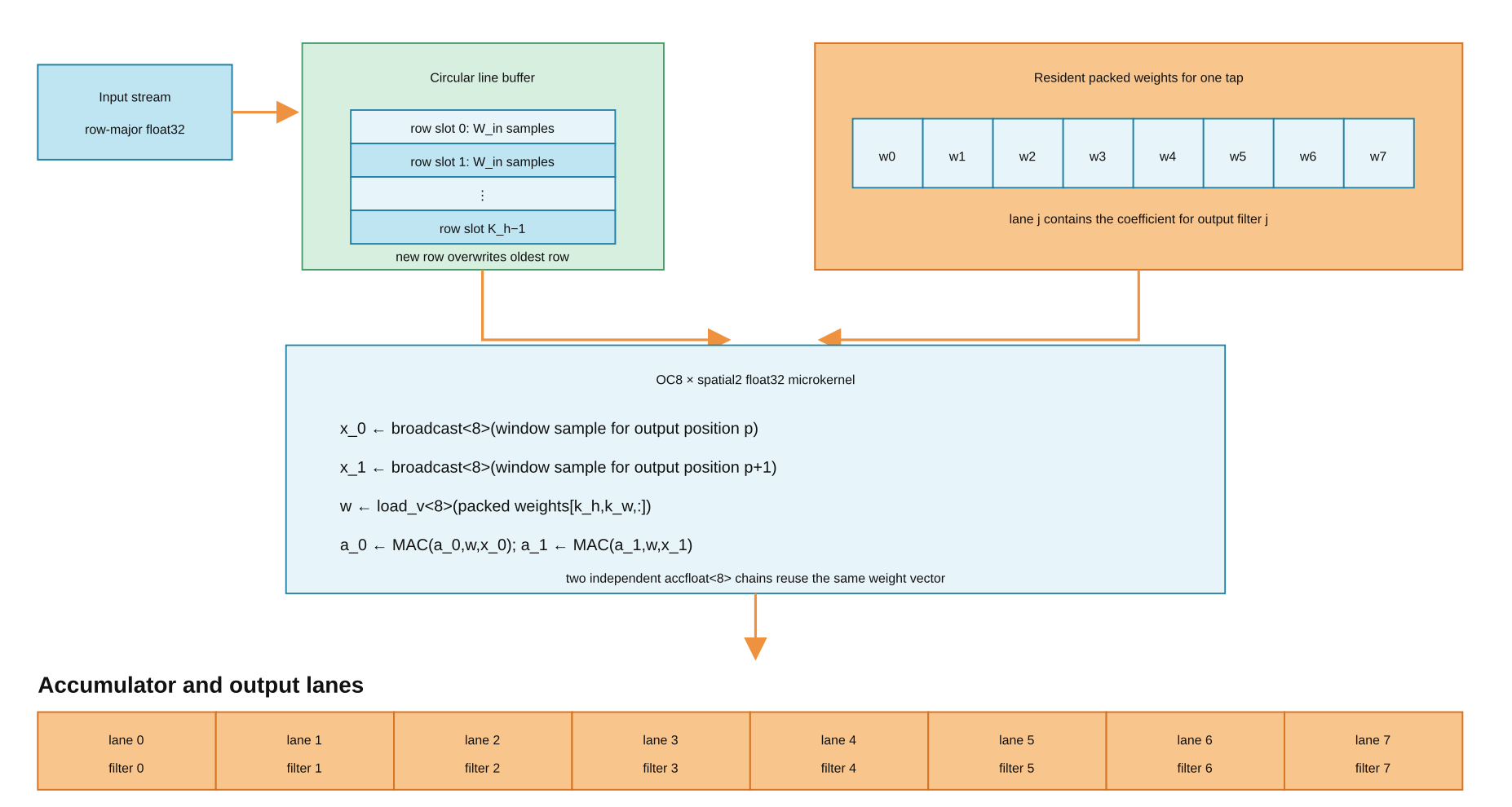}
    \caption{Circular line-buffer and vector-register mapping of the float32 Conv2D kernel. Each input sample is broadcast across eight lanes, while the corresponding packed weight vector contains coefficients for up to eight output filters. Two neighbouring spatial positions reuse the same weight vector}
    \label{fig:CNNImage}
\end{figure}

The principal vectorisation axis is the output-channel dimensions, as illustrated in Fig. \ref{fig:CNNImage}. Weights are packed offline in the order ([$K_h$][$K_w$][$C_{\mathrm{in}}$][$C_{\mathrm{out,pad}}$]), where ($C_{\mathrm{out,pad}}$) is the output-channel count rounded up to a multiple of eight. For each kernel tap, one scalar input sample is broadcast across eight vector lanes, while weight lane (j) contains the coefficient for output filter (j). A single vector MAC therefore advances eight output filters concurrently. Unused output-channel lanes contain zero-values weights and biases. 

Two neighbouring output positions are processed using independent 8-lane accumulator chains. The same packed weight vector is reused with two broadcast input samples, providing parallelism over both output channels and neighbouring spatial positions. After all ($K_h, K_w$)taps have been accumulated, the results are converted to float32 vectors. Biases initialise the accumulators, while an optional ReLU is implemented as a vector maximum before the output is written. An odd final output column uses a single accumulator. Since loop bounds, tail handling and padding behaviour are determined at compile time, the arithmetic instruction sequence does not depend on the input values.

Weights, biases and the circular line buffer reside in tile-local memory. If a layer cannot be represented efficiently by one worker, the implementation can partition the operation by output-channel range or output-row range. Every spatial worker maintains its own line buffer and receives the required input rows through AIE stream multicast. The resulting row partitions are restored by an ordered binary tree. This avoids explicit halo-row exchange, at the cost of replicated input delivery. Input-channel partials can alternatively be evaluated by separate workers and combined through binary AIE sum kernels, with bias and activation applied only after the final reduction. These strategies provide parallel line-buffering without requiring intermediate processing in programmable logic.

\subsection{Dense kernel}

Traditionally ML algorithms implement dense layers towards the end to provide the single prediction. Dense layers apply an affine transformation to an input feature vector, followed by an optional element-wise activation:

\begin{equation} \label{eq:DenseEq1}
    y_j = \sigma\left(b_j + \sum_{i=0}^{N_{\mathrm{in}}-1} W_{j,i} x_i\right), \qquad \mathbf{y} = \sigma\left(W\mathbf{x} + \mathbf{b}\right),
\end{equation}

where ($N_{\mathrm{in}}$) and ($N_{\mathrm{out}}$) denote the input and output dimensions, respectively. The kernel uses float32 inputs, weights, biases and outputs and maps the matrix-vector product onto the native AIE matrix-multiplication unit.

The implementation uses the native float matrix-multiplier operation, which is 8b-aligned and hence the input and output dimensions are therefore padded internally as

\begin{equation} \label{eq:DenseEq2}
    2\left(\lceil\frac{N_{\mathrm{out}}}{2}\right)\rceil.
\end{equation}

Padding values and their corresponding weights are set to zero and do not modify the mathematical output.

For each input vector, the ($N_{\mathrm{in}}$) logical features are first received from the stream and stored in an aligned tile-local cache. This cache is reused for every pair of output neurons, avoiding repeated input-stream transfers. Weights are transformed offline from their mathematical ($N_{\mathrm{out}}\times N_{\mathrm{in}}$) representation into contiguous ($8\times 2$) tiles: 

\begin{equation} \label{eq:DenseEq3}
    W_{\mathrm{packed}} \left[\left\lceil N_{\mathrm{out}}/2\right\rceil\right]\left[\left\lceil N_{\mathrm{in}}/8\right\rceil\right][8][2].
\end{equation}

Each aligned weight load therefore contains the coefficients connecting eight adjacent input features to two adjacent output neurons.

\begin{figure}[h]
    \centering
    \includegraphics[width=0.75\linewidth]{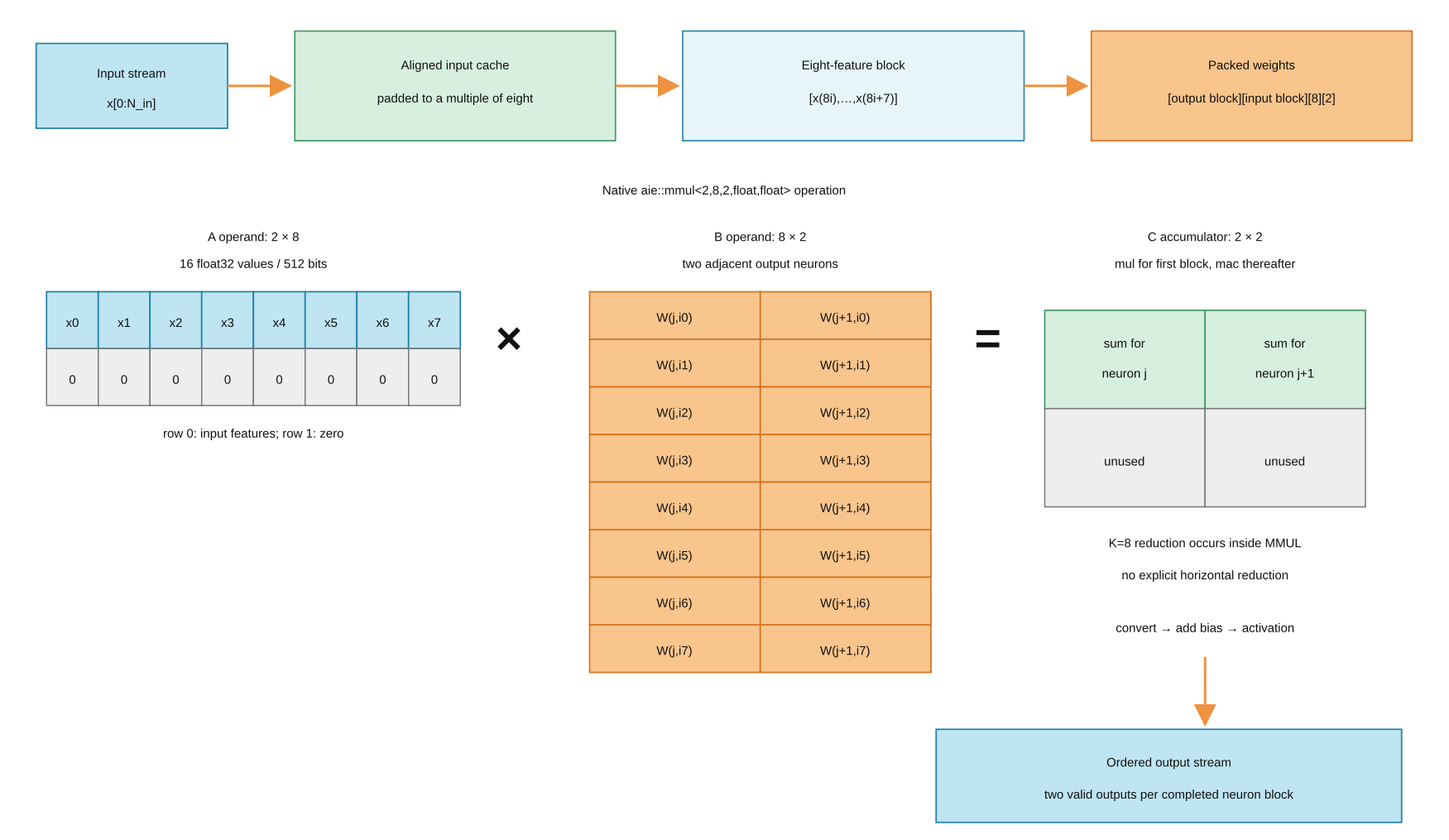}
    \caption{Operand and accumulator mapping of the float32 Dense kernel onto the native $2\times8\times2$ AIE matrix multiplication operation. Row zero of the input operand contains eight features, while row one is zero-filled. Each operation advances the dot product of two adjacent output neurons.}
    \label{fig:DenseImage}
\end{figure}

The register mapping is illustrated in Fig. \ref{fig:DenseImage}. Row zero of the ($2\times 8$) input operand contains eight input features, while the second row is filled with zeros. The ($8\times 2$) weight operand contains the coefficients for two output neurons. The resulting ($2\times 2$) accumulator therefore contains two useful partial sums in its first row; the second row corresponds to the zero-values input row and is discarded. The first input block initialises the accumulator using a matrix multiplication, while subsequent blocks use multiply-accumulate operations:

\begin{equation} \label{eq:DenseEq4}
    \begin{aligned}
        C^{(0)} &= A^{(0)}B^{(0)},\\
        C^{(k)} &= C^{(k-1)} + A^{(k)}B^{(k)}.
    \end{aligned}
\end{equation}

The reduction over each group of eight input features is performed internally by the matrix multiplication operation, so no explicit horizontal vector reduction is required.

\begin{figure}[h]
    \centering
    \includegraphics[width=0.75\linewidth]{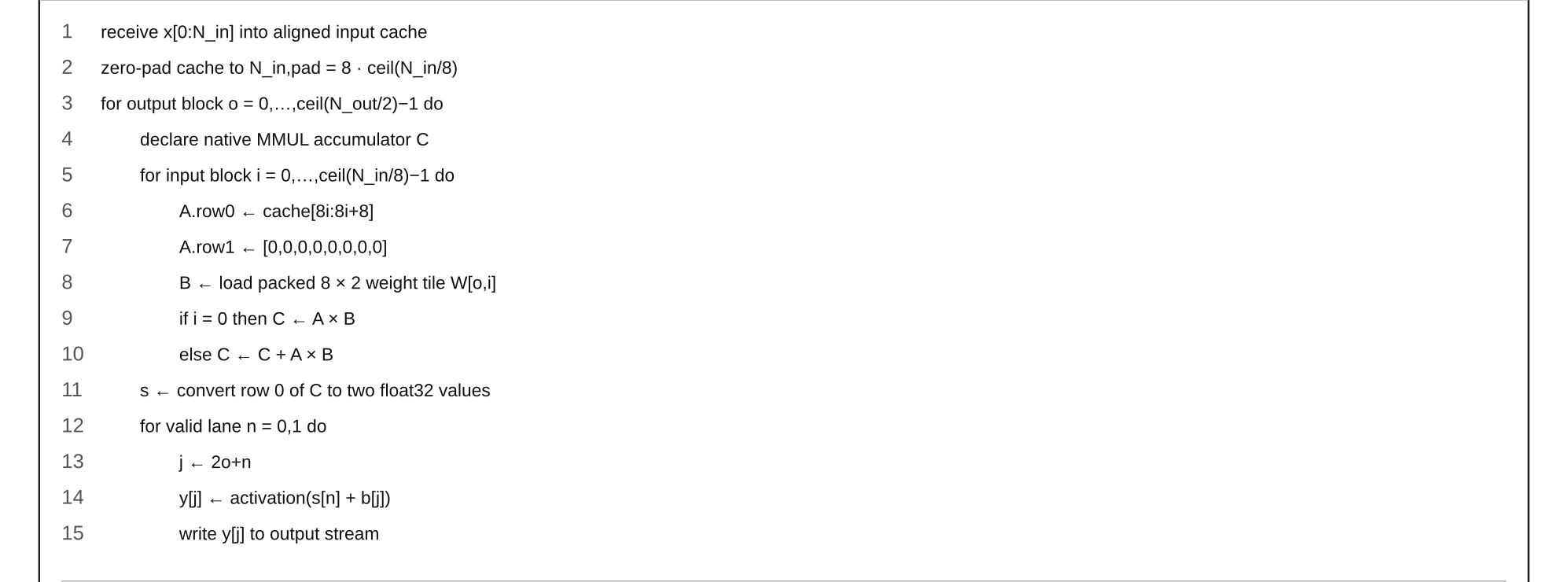}
    \caption{Buffered execution of the float32 Dense kernel. The aligned input vector is cached once and reused while native matrix multiplication operations evaluate pairs of output neurons over eight-feature blocks.}
    \label{fig:DensePseudocode}
\end{figure}

After all input blocks have been processed, the two valid accumulator values are converted to float32. The corresponding biases are added, the optional activation is applied, and the results are written to the output stream in neuron order. ReLU and sigmoid activations can be applied after bias addition, while whole-vector transformations such as softmax require a separate kernel. The input dimensions, output dimensions, padding and activation are compile-time properties; consequently, input values do not alter the matrix-multiplication loop counts or arithmetic instruction sequence. The full execution order is summarised in Fig. \ref{fig:DensePseudocode}.

Weights and biases are stored as aligned constants in tile-local memory. A single worker processes its assigned output neurons sequentially in pairs while retaining only one matrix multiplication accumulator. When the complete weight matrix cannot be assigned efficiently to one tile, the output dimension can be divided into contiguous neuron ranges. Each worker receives the complete input vector but stores only the weights and biases for its output range. The independent output streams can then be concatenated in neuron order through AIE stream-connected stitch kernels. The current generalised implementation does not divide individual dot products across tiles; such an input-dimension partition would additionally require partial-sum reduction before bias and activation.


\section{Validation Examples for kernels}

The reference CNN network follows the Global Trigger jet-image classifier~\cite{LargeRJetCNN}. It's input is a single-channel image of calorimeter trigger-tower transverse energies centred on a seed jet. In this work, all validation studies use the $18\times18$ image configuration. The preprocessing follows the published model: zero-suppressed pixel energies transform as $x\mapsto\log(1+x)$ and the resulting image is normalised to unit total energy. These operations must be applied identically before the Keras, quantised, and AIE inferences so that the comparison isolates the numerical implementation rather than differences in input preparation. The network architecture can be seen in Tab. \ref{tab:CNNArchitecture}.

\begin{figure}[h]
  \centering
  \begin{tikzpicture}[
      node distance=3.5mm,
      block/.style={draw, rounded corners, align=center, minimum height=8mm,
                    inner xsep=3mm, font=\scriptsize},
      flow/.style={-{Latex[length=1.8mm]}, thick}
    ]
    \node[block] (input) {Input\\$18\!\times\!18\!\times\!1$};
    \node[block, right=of input] (conv1) {Conv2D\\$7\!\times\!7$, $s=2$\\5 filters\\$6\!\times\!6\!\times\!5$};
    \node[block, right=of conv1] (conv2) {Conv2D\\3 filters\\$6\!\times\!6\!\times\!3$};
    \node[block, right=of conv2] (conv3) {Conv2D\\1 filter\\$6\!\times\!6\!\times\!1$};
    \node[block, below=6mm of conv3] (flat) {Flatten\\36};
    \node[block, left=of flat] (dense1) {Dense\\16};
    \node[block, left=of dense1] (dense2) {Dense\\2 outputs};

    \draw[flow] (input) -- (conv1);
    \draw[flow] (conv1) -- (conv2);
    \draw[flow] (conv2) -- (conv3);
    \draw[flow] (conv3) -- (flat);
    \draw[flow] (flat) -- (dense1);
    \draw[flow] (dense1) -- (dense2);
  \end{tikzpicture}
  \caption{Architecture of the $18\times18$ jet-image CNN implemented on the
  AI Engines. Tensor dimensions exclude the batch dimension.}
  \label{tab:CNNArchitecture}
\end{figure}

The AIE implementation retains float32 weights, activations, biases, and accumulations throughout. It therefore implements the floating-point reference model rather than the 8b quantised FPGA arithmetic.

The published FPGA design uses quantisation-aware training with 8b arithmetic and targets the AMD Versal VP1802 and VP2802 devices at 200MHz. For the $18\times18$ configuration, a latency of 133 cycles ($0.67\,\mu\mathrm{s}$), an initiation interval of 36 cycles, and utilisation of approximately 8\% of the LUTs and 3\% of the Flip-Flops in one super logic region are reported, with negligible BRAM and DSP use. The same source reports an AUC of 0.88 for the full-precision $18\times18$ model and 0.877 for the 8b quantised FPGA configuration. 

For this reason the AIE results discussed below will focus on the VP2802 FPGA as well since this is the package that contains the AIEs in the Versal family targeted by projects such as the ATLAS Global Trigger.

\section{Results}

The complete BDT and CNN implementations are evaluated using float32 weights, parameters, activations, and accumulation. The CNN study uses 100 large-R jets generated with Delphes, mimicking ATLAS, in the $18 \times 18$ image format, while the BDT uses synthetic forests and pseudo data because there is no available open dataset for the target model. The CNN sample is used for event-by-event numerical validation rather than a new measurement of physics performance, as the latter is being discussed in Ref.~\cite{LargeRJetCNN}. The BDT study similarly establishes inference correctness and model capacity, not the discrimination performance of a particular trained forest. Unless stated otherwise, timing values are from cycle-accurate AIESim and resource values are obtained using Vitis 2024.2 post-placement reports, rather than measurements on physical hardware.

\subsection{BDT performance}

\begin{figure}[h]
    \centering
    \begin{subfigure}[t]{0.49\textwidth}
        \centering
        \includegraphics[width=\linewidth]{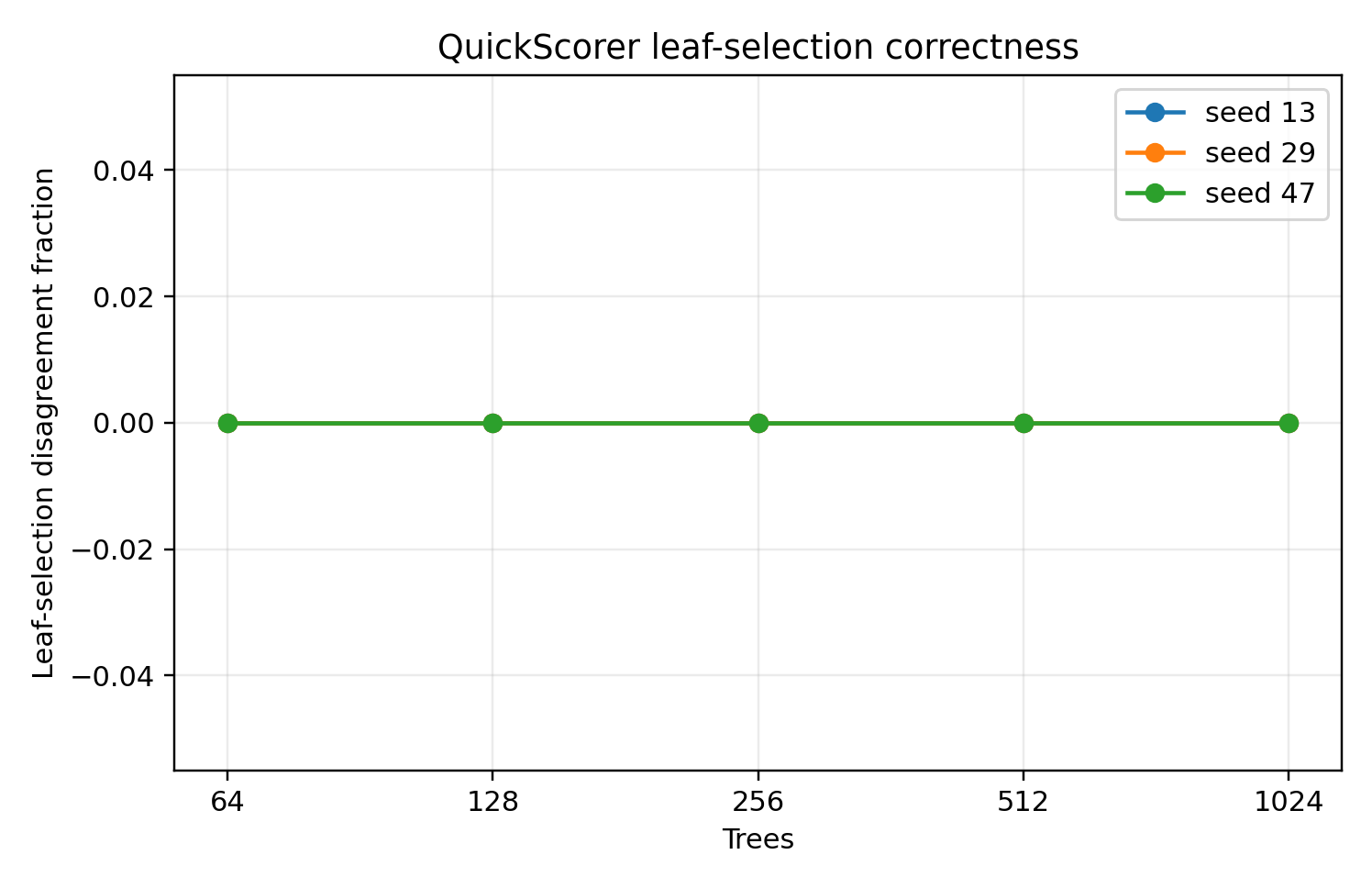}
        \caption{Leaf-selection disagreement fraction.}
    \end{subfigure}\hfill
    \begin{subfigure}[t]{0.49\textwidth}
        \centering
        \includegraphics[width=\linewidth]{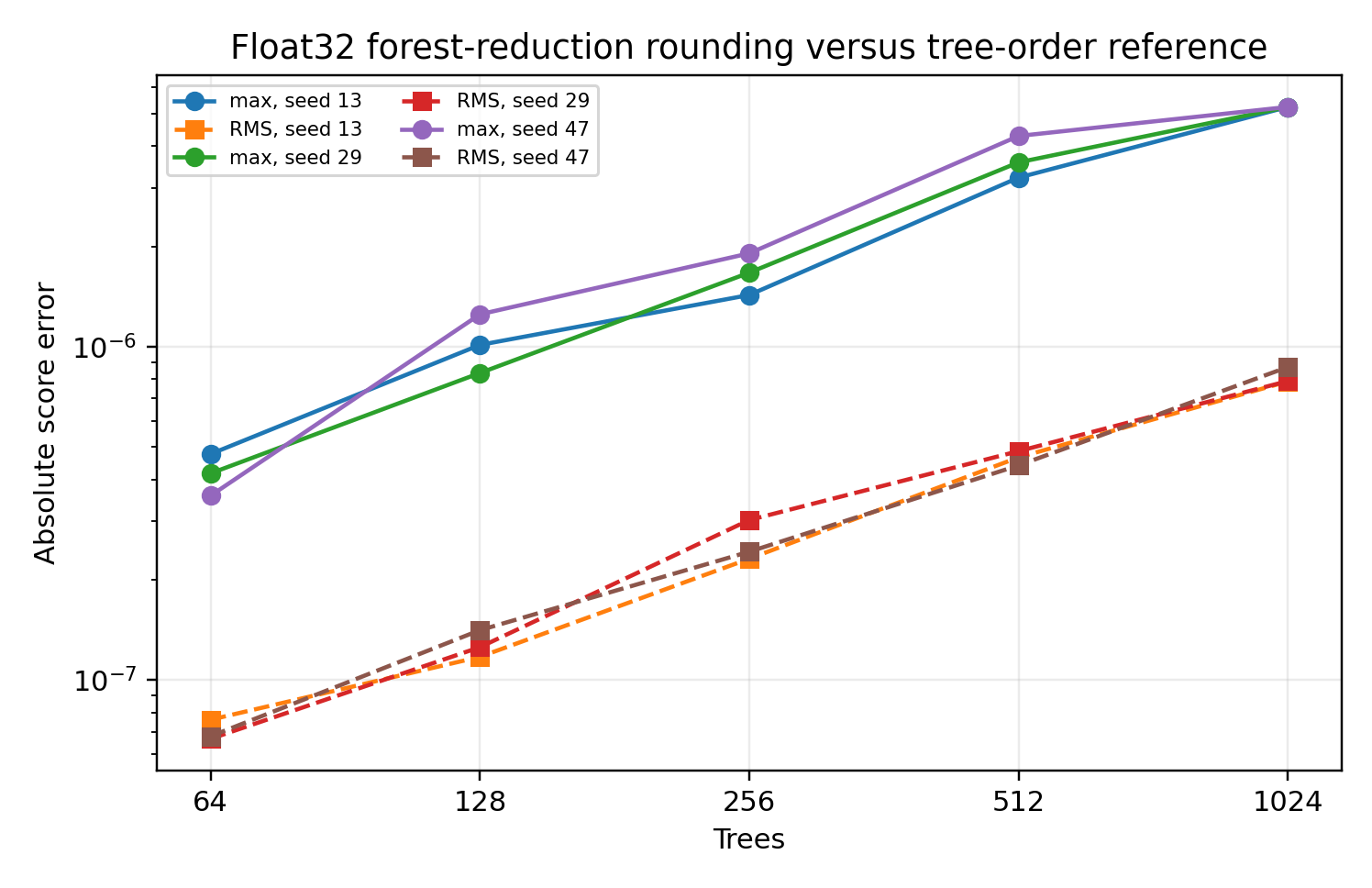}
        \caption{Float32 reduction-order error.}
    \end{subfigure}
    \caption{BDT correctness for depth-five forests. Score differences are software emulations of the multi-worker reduction order and do not hide any branch or leaf disagreement.}
    \label{fig:BDTCorrectness}
\end{figure}

The BDT implementation was tested with balanced, depth-five regression forests containing 64--1024 trees, 32 input features, and three independent random seeds per size. Each split was tested immediately below, at, and above its float32 threshold, in addition to 4096 random inputs per model and special values including signed zero, infinities, and NaNs. Across 35,408,448 tree-event comparisons, the BDT masks and an independent scalar traversal, executed in CPU with XGBoost, selected the same leaf in every case. With identical tree-order accumulation, the final scores were bitwise identical. The small errors in Fig.~\ref{fig:BDTCorrectness} arise only when the per-worker partial sums are combined in the AIE reduction-tree order. Even for 1024 trees the largest emulated difference is $5.25\times10^{-6}$ and the largest RMS difference is $8.70\times10^{-7}$.

The representative 64-tree forest contains 4032 nodes and 2048 leaves. Its AIE graph uses eight forest workers, seven feature-broadcast kernels and seven score-reduction kernels, for 22x kernels in total. Running AIESim on 100 events produces a mean steady-state interval of

\begin{equation} \label{eq:BDTInterval}
    \tilde{\tau}_{\text{interval}} = 2.133 \pm 2.38\times10^{-7}\, \mu\mathrm{s},
\end{equation}

where the uncertainty quoted corresponds to the RMS over the 100 events. The developed kernel is pipelined, meaning that the first output of the pipeline will be emitted at $\tau_{\text{first}} = 3.9\, \mu\mathrm{s}$, whereas all the next at the $\tau_{\text{interval}}$ value shown in Eq. \ref{eq:BDTInterval}, which both of them are below the fixed latency constraint of $<10\, \mu\mathrm{s}$.

The kernel has also been tested with more trees in order to probe multi-tile performance in order to identify routing constraints. For that purpose, 128, 256, 512 and 1024 trees have been generated, increasing the number of forest workers from 10 to 74, while assigning at most 14x trees to each worker. Till 256 trees the latency increased linearly, whereas at 512 a plateau was identifies at $\tau_{\text{interval max}} = 7.128\, \mu \mathrm{s}$. The fixed latency for a larger number of trees is due to the compute bound of the kernel driving the result, as the maximum latency of 14x trees per tile was hit, hence any addition of workers makes the result dominated by the slowest worker. The 14x tree limit was obtained by optimising over the memory utilisation of the tile. For this reason, 14x trees per tile with depth-five correspond to 29~kB of memory utilisation from a total of 32~kB available in every tile from v1.0 generation.

For context, BDTs developed for trigger applications~\cite{ATLASBDT} report sizes of 77 depth-six trees and 1381 nodes, using 10b split values and fixed-point leaf values. Such trees are extremely cheap in FPGA as they don't utilise many resources. Tiny BDT models like that are more efficient in FPGAs rather than AIEs. However, if float32 precision is requested and a larger amount of nodes, then the AIE kernel enabled this extra face space previously inaccessible via the FPGA-only implementations.

\subsection{CNN performance}

The CNN comprises three Conv2D/ReLU stages followed by flatten, a 16-output Dense/ReLU layer, and two-output Dense/Softmax layers, for 1042 trainable parameters. The complete float32 model was compared with the original Keras model, a converted float32 ONNX model, and supplied 8b quantisation-aware trained (QAT) predictions. Both output nodes were retained, yielding 200 values. ONNX agrees with Keras to a maximum absolute different of $5.36\times10^{-7}$. The AIE output agrees with ONNX to a maximum absolute difference of $1.75\times10^{-4}$, with mean absolute and RMS differences of $4.21\times10^{-5}$ and $6.66\times10^{-5}$, respectively. The dominant difference is localised to the approximate fused Softmax lookup-table stage, whereas all preceding Conv2D, Flatten and Dense/ReLU prefixes have maximum errors below $1.91\times10^{-6}$.

\begin{figure}[h]
    \centering
    \includegraphics[width=\textwidth]{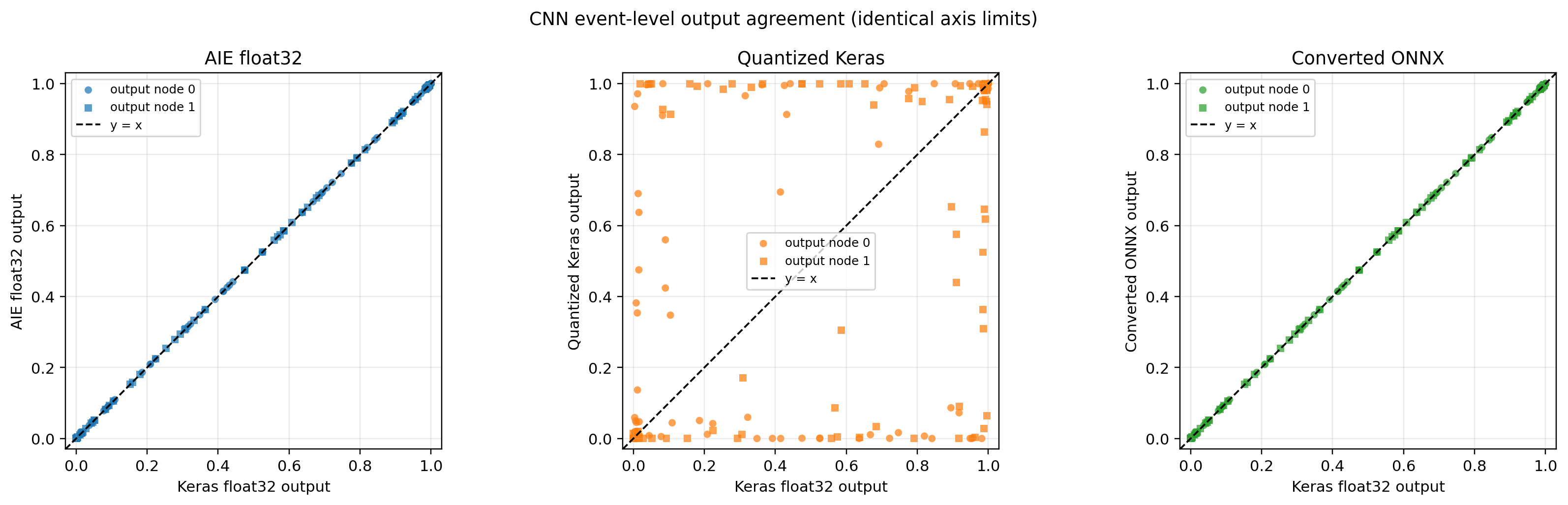}
    \caption{Event-level comparison of both CNN output nodes with the Keras float32 reference. All panels use identical axes. The AIE model uses float32 throughout, the QAT predicitions are a software-model comparison and not an int8 AIE implementation.}
    \label{fig:CNNOutputComparison}
\end{figure}

As shown in Fig.~\ref{fig:CNNOutputComparison}, the float32 AIE and ONNX outputs track the Keras reference over the full output range. The QAT predictions differ more substantially, with a mean absolute difference of 0.273 and an RMS difference of 0.428 relative to Keras. For this sample, the AIE-to-ONNX mean absolute difference is $1.54\times10^{-4}$ of the QAT-to-float difference. This comparison motivates that the AIE version of the CNN is essentially the same as the Keras float32 model quoted in the Ref.~\cite{LargeRJetCNN} note, and hence the AUC values shown can be reached if the models were supplied with identical inputs. 

The complete graph occupies 14x compute tiles and 14x kernels after placement (1x kernel/tile). Summed program memory is 35~kB, with a maximum of 4~kB/16~kB (28.3\%) on one core. Compiler-reserved data memory is 82~kB across the graph, with a maximum memory-group occupancy of 14~kB/32~kB (44.4\%). Eight DMA FIFO allocations use 3.9~kB, and the graph contains 15 inter-kernel streams and one 32b input and output PLIO at 312.25~MHz. No placement warnings, waivers, or memory oversubscription were reported. The first convolution is spatially split over three parallel line-buffer kernels and reassembled by two height-stitch kernels. This demonstrates a compiled multi-buffer implementation which enabled the target of $10\, \mu\mathrm{s}$ to be met.

The 100-event AIESim run produced the first complete two-score output after $\tau_{\text{first}} = 9.39\, \mu\mathrm{s}$. The 99 consecutive output intervals of the pipelined design have a mean and a sample standard deviation of

\begin{equation}
    \tilde{\tau}_{\text{interval}} = 2.025 \pm 0.006\, \mu\mathrm{s}.
\end{equation}

\begin{figure}[h]
    \centering
    \includegraphics[width=\textwidth]{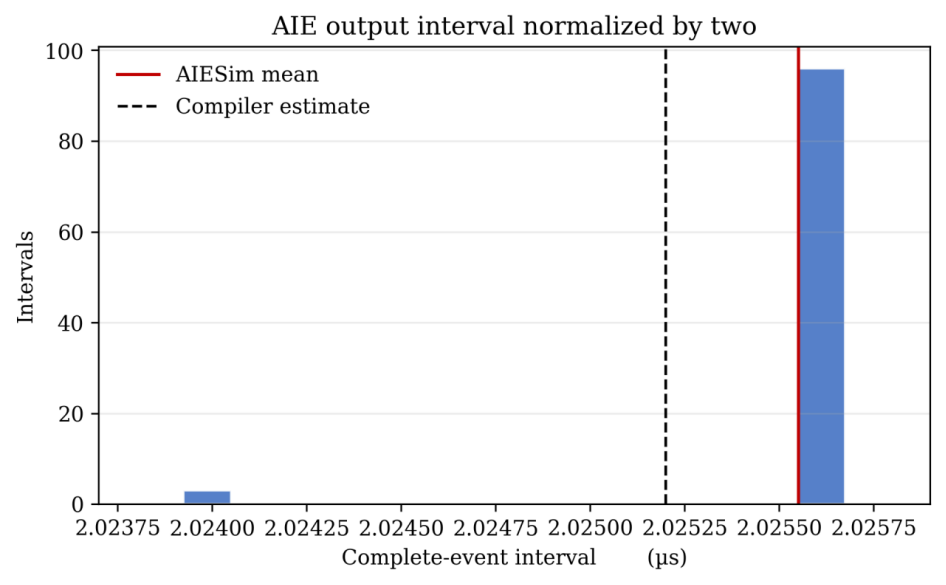}
    \caption{Distribution of sequential complete-event output intervals for the 100-event CNN AIESim run. The dashed and solid lines show the compiler estimate and AIESim mean, respectively.}
    \label{fig:CNNVariation}
\end{figure}

The variation of the output latencies can be seen in Fig. \ref{fig:CNNVariation}

The FPGA CNN uses an 8b QAT implementation of the same $18\times18$ topology and reports $0.67\, \mu\mathrm{s}$ latency and a $0.18\, \mu\mathrm{s}$ initiation interval at 200~MHz. This study shows that both approaches are meeting the latency requirement; however,owing also for larger CNN architectures which are prohibited for FPGAs.

\begin{figure}[h]
    \centering
    \includegraphics[width=\textwidth]{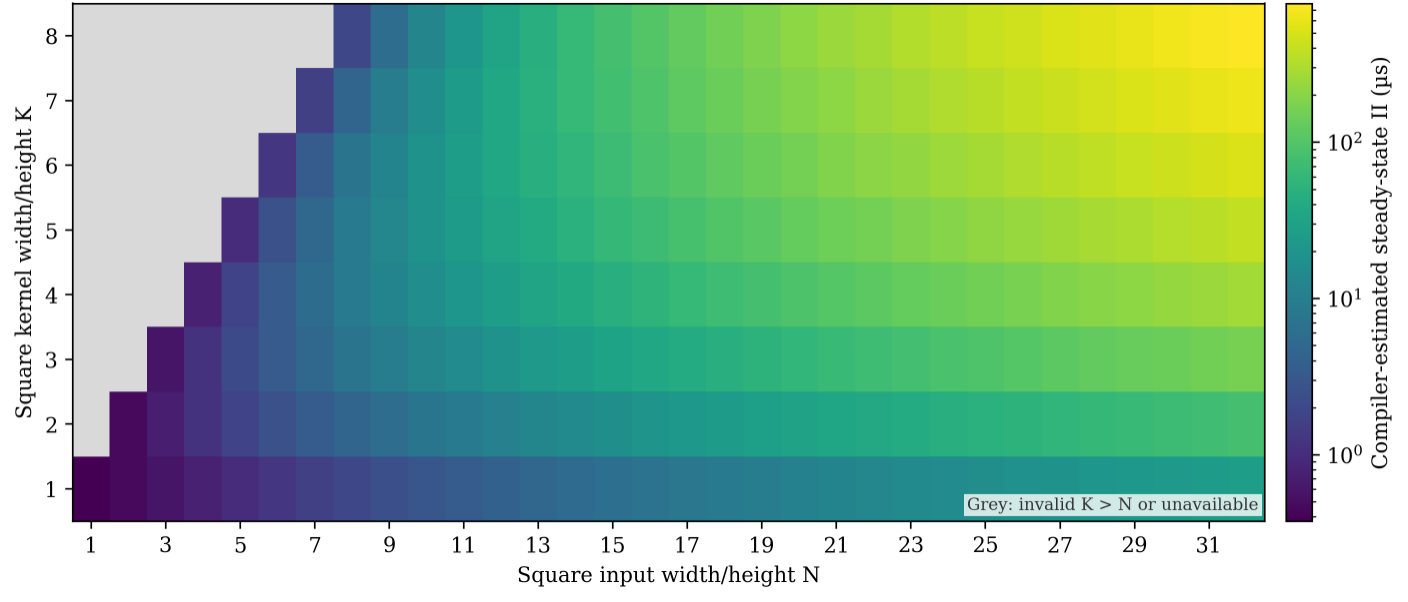}
    \caption{Heat map on latencies for single tile implementation of the Conv2D layer in float32 precision. No tiling has been applied as the following heatmap is used as guide so that tiling in realistic models can be applied and succeed the objective of lower latency.}
    \label{fig:CNNHeatMap}
\end{figure}

The largest part of the latency for the Convolution network originates from the first convolution due to the first-come first-served style implementation, for this reason the convolutional kernels has been scanned for different configurations to identify the latency that would inflict on the overall performance. In Fig. \ref{fig:CNNHeatMap} there is a clear region where the I/O latency for the data stream is making the performance of the kernel to be overtaken over the compute and hence applications like unless tiled at the input layer they would not be able to implement in a single tile with the given latency requirement.

\section{Conclusion/Discussion}

These results demonstrate complete float32 BDT and CNN inference in cycle-accurate AIESim which meet the latency requirements for fixed latency triggers. The purpose of the study is not to replace deeply pipelined shallow FPGA oriented models but rather to show that AIEs provide a viable intermediate implementation target for larger or floating-point models under fixed-latency constraints. Physics-board timing, power, and a controller precision sweep remain outside the scope of this work. Table~\ref{tab:AIEFPGAComparison} summarises the demonstrated AIE results and the published FPGA baselines. 

\begin{table}[h]
    \centering
    \begin{tabular}{c c c c c c}
         \hline
         \textbf{Model} & \textbf{Impl.} & \textbf{Arithmetic} & \textbf{First Output} & \textbf{II} & \textbf{Resources} \\
         \hline
         CNN & AIE & float32 & 5.39 $\mu$s & 2.025 $\mu$s & 14 tiles/kernels; PM 35.6~kB \\
         CNN & FPGA & fixed8 & 0.67 $\mu$s & 0.18 $\mu$s & 8\% LUT, 3\% FF \\
         BDT & AIE & float32 & 3.9 $\mu$s & 2.13 $\mu$s & 22 tiles/kernels; PM 11.2~kB \\
         BDT & FPGA & fixed3 & 0.011 $\mu$s & Not Reported & 5376 LUT, 4908 FF \\
         \hline
    \end{tabular}
    \caption{AIE demonstrations and representative FPGA baselines. Resource units are architecture-specific and are deliberatly not converted. Both FPGA models were designed with FPGA deployment in mind and hence compute was reduced by nature.}
    \label{tab:AIEFPGAComparison}
\end{table}

\ack{This work has been partially funded by the Eric \& Wendy Schmidt Fund for Strategic Innovation through the CERN Next Generation Triggers project under grant agreement number SIF-2023-004. We also, gratefully acknowledge the support of the UK's Science and Technology Facilities Council (STFC). NCH is supported by the STFC UCL Centre for Doctoral Training in Data Intensive Science  (ST/W00674X/1) and by UCL and industry funds. DWM is supported by the National Science Foundation under Grant No. PHY-2310094.}

\data{No new data were created or analysed in this study.}

\funding{
The contributors of this paper are supported by:
\begin{enumerate}
    \item Eric \& Wendy Schmidt Fund for Strategic Innovation (SIF-2023-004)
    \item Science and Technology Facilities Council (STFC) (ST/W00674X/1)
    \item National Science Foundation (PHY-2310094)
\end{enumerate}}

\bibliographystyle{iopart-num}
\bibliography{references_iop_robust}

\end{document}